# Optical pumping and initialization of a hole spin in site-controlled InGaAs pyramidal quantum dots

R. A. Barcan[1,2], I. Samaras[1], K. Barr[1], G. Juska[3], E. Pelucchi[3] & K. G. Lagoudakis[1†]

[1] *Department of Physics, University of Strathclyde, Glasgow G4 0NG, United Kingdom*
[2] *Sussex Centre for Quantum Technologies, University of Sussex, Brighton BN1 9QH, United Kingdom*
[3] *Tyndall National Institute, University College Cork, Cork T12 R5CP, Ireland*

We investigate site-controlled $In_{0.25}Ga_{0.75}As$ quantum dots in (111)B GaAs pyramidal recesses as spin qubits. Combining scanning confocal cryomicroscopy, magneto-photoluminescence studies and resonant excitation, we identify and isolate a positively charged exciton with a hole-spin in its ground state. Application of a strong 5 T magnetic field parallel to the growth axis, induces a fourfold splitting of the energy levels of the positively charged exciton creating an optically addressable double-lambda system. We combine weak above-band and resonant excitation to demonstrate spin pumping and high-fidelity spin initialization through all four optical transitions and study the system behavior as a function of the resonant driving strength showing the existence of a robust spin that can be optically pumped and initialized. These results demonstrate the potential of these quantum dots for precise spin manipulation and their relevance for future quantum hardware.

## I. Introduction

Semiconductor quantum dots (QDs) have emerged as promising platforms for quantum technologies, primarily due to their atom-like properties and the ability to tailor their electronic and optical characteristics through quantum confinement effects that can be fine-tuned according to the various growth techniques [1],[2]. Epitaxially grown quantum dots exhibit high stability, brightness, and emission tunability, making them ideal candidates for applications such as spin-based qubits, single- and entangled- photon sources [3]-[5].

Unlike self-assembled QDs, which exhibit randomness in positioning and spectral properties, site-controlled QDs allow for precise placement within photonic structures, enabling enhanced light-matter interactions and offer the potential for integration into scalable quantum networks [6]-[7]. Among the various platforms of deterministically positioned QDs [8]-[17], the platform of site-controlled InGaAs quantum dots grown by Metal-Organic Vapor Phase Epitaxy (MOVPE) in pyramidal recesses on (111)B GaAs substrates stands out. These quantum dots exhibit high spatial uniformity [18] and thanks to the high symmetry ($C_{3v}$), they have reduced fine-structure splitting (FSS) enabling the generation of polarization-entangled photons [19]. Another remarkable property of these dots is that they often show charged excitonic complexes in their spectral signatures suggesting that they naturally contain trapped electron or hole spins enabling the study of magneto-photoluminescence properties [20],[21]. The combination of all these properties make them particularly attractive for quantum optics and quantum information processing applications.

Here we identify a positively charged quantum dot and after application of a magnetic field in the Faraday configuration, we show that thanks to the Zeeman splitting of both hole-spin ground state and trion state, a double lambda-system with distinct transition energies is created in the quantum dot. Using resonant optical driving and photon detection on the two distinct transitions of each lambda-level structure, we investigate spin pumping and initialization for each hole-spin ground state.

## II. Materials and methods
### A. Site-controlled InGaAs pyramidal QDs

The study is conducted on a sample comprising $In_{0.25}Ga_{0.75}As$ QDs with a nominal thickness of 0.85 nm, grown on a (111)B GaAs substrate. The substrate is pre-patterned with tetrahedral pyramidal recesses with a 7.5 μm pitch to ensure deterministic positioning. This precise and flexible patterning technique guaranteed accurate positioning and size uniformity of the QDs, along with high spectral uniformity and narrow linewidths that are key factors for their application in quantum hardware [18]. The sample(s) were grown by MOVPE with standard conditions, whose description can be found elsewhere [22].

### B. Magneto-optic cryo-microscopy setup

Throughout the experiment, the sample is placed in a magnetic cryostat (Oxford Microstat-MO) combined with an integrated helium cryo-recirculator (ColdEdge Stinger) for closed-cycle operation [23]. The system maintains the sample at a cryogenic temperature of approximately 6.8 K. Additionally, the cryostat supports a homogeneous magnetic field of up to 5 T, aligned along the central axis of the sample chamber, enabling the observation of the magneto-optical properties of the system studied. For the analysis of the QD's magneto-optic properties, the magnetic field is incrementally increased by 0.5 T up to a maximum of 5 T. At each step, the intensity and energy of the PL emissions are recorded. Optimization of the collected photons is performed at each magnetic field by finely tuning the positioning of the excitation/collection optics to correct for any drift or misalignment induced by the magnetic field forces on the optical system. The magnetic

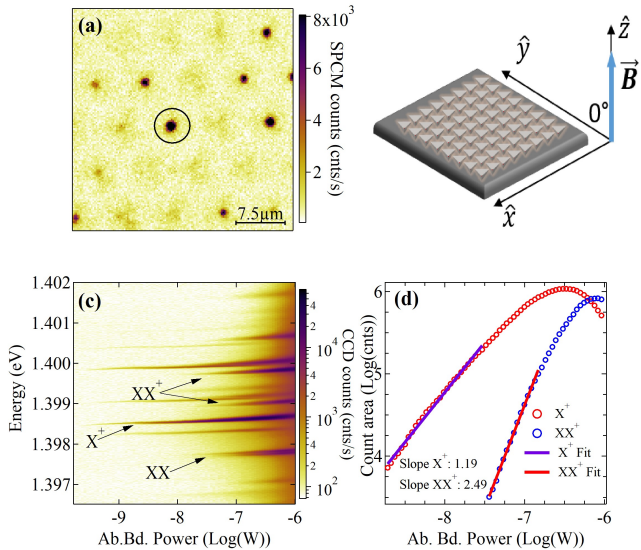

FIG. 1: (a) 2D photoluminescence map of the scanned QD array, with a color scale indicating emission intensity in SPCM counts per second. The circle marks the specific QD that was further investigated in this work. (b) Schematic illustration of the array of pyramidal recesses with the QDs placed at the tip of the pyramid, showing the orientation of the applied magnetic field. (c) Above band excitation power dependence of the emission spectrum of the studied QD. The charged exciton is marked by $X^+$ while the biexciton and charged biexciton are marked by XX and $XX^+$, respectively. (d) Variation of the Lorentzian emission area with excitation power for the trion and the charged biexciton showing a near-linear and quadratic power dependence

field is applied in the Faraday configuration (parallel to the growth axis), but due to the inherent symmetry of the QDs, the magnetic field effects resemble those of an oblique configuration for self-assembled QDs [20],[24]-[26]. As a result, we quote effective g-factors for the electron and hole and we use the subscript θ for the hole and trion states. Optical excitation and photon collection is enabled via a 0.5 N.A. long working-distance (~12 mm) microscope objective lens (Mitutoyo, 100x NIR Plan Apo).

### C. Two-dimensional mapping and spectral analysis

Prior to focusing on a specific QD, a surface map of the sample is generated using a custom-built scanning confocal microscopy setup. This allows for the quick characterization of large numbers of QDs both in spectrally integrated and spectrally resolved modes. For spectroscopic analysis, the photoluminescence (PL) signal is filtered through a custom-built double Czerny-Turner spectrometer (2 x 1 m), providing a spectral resolution of approximately 8 μeV. The spectrally resolved signal can be directed either to a CCD array for the acquisition of the full spectra or to a single-photon counting module (SPCM) for photon count rate measurements at specific energies.

### D. Above-band and resonant excitation

Continuous-wave above-band laser excitation is used in a two-fold manner. It allows for PL measurements of the QD which permits the characterization of the system under magnetic field and is additionally used under low excitation power for the randomization of the ground spin states during the spin pumping and initialization experiments. To address individual transitions of the QD under magnetic field for the spin pumping and initialization experiments, a CW single-mode Ti:Sapphire laser with ~40GHz modehop-free scanning is used to resonantly drive the target transitions. The resonant laser frequency is monitored using a Fizeau-type wavemeter with a relative frequency resolution of 10 MHz, ensuring precise real-time frequency monitoring and control for the desired transitions. The resonant laser power and polarization is set using a combination of linear polarizers combined with a quarter-wave and a half-wave plate.

## III. Results and discussion
### A. Spectral lines and above band excitation power dependence

FIG. 1(a) shows a map of a spectrally integrated PL confocal scan of a region on the sample where the QD studied here is located. Several QDs within the scan area show high relative counts allowing for the identification of the triangular pattern of the pyramidal recesses containing individual QDs, similar to what is schematically presented in FIG. 1(b). We focus on a QD that shows charged excitonic complex signatures, indicated by the circle in the PL scan map.

The PL power-dependence measurements under spectral resolution shown in FIG. 1(c), reveal a rich excitonic complex behavior. The energetic ordering of the peaks allows one to associate the emission of certain peaks to the positively charged exciton $X^+$, and biexcitons $XX^+$ and XX, indicated with white arrows. A comparison of the PL intensity behavior of the charged exciton $X^+$ and biexciton $XX^+$ shown in FIG. 1(d), clearly shows the linear vs quadratic behavior expected for this family of excitonic complexes. The $X^+$ emission is observed to saturate at approximately $P_{Ab.Bd}^{sat} = 250\,nW$. A common feature in all spectral lines is a slight blue-shift in emission energy for increasing above-band laser power. This blue-shift is likely a consequence of Coulomb interactions involving charge carriers present in the surrounding nanostructures, such as lateral wires [27].

### B. Magnetophotoluminescence and Zeeman splitting

To access the hole-spin levels of the positively charged QD, a magnetic field is applied in the Faraday configuration and the PL spectrum of the charged exciton is recorded in two opposite circular polarizations, showing a fourfold splitting. The behavior of the splitting is a combination of the linear-in-field Zeeman interaction between the particle's magnetic moment and the external magnetic field, and a quadratic contribution due to the diamagnetic shift. Fitting the split peaks as a function of the magnetic field for the two circular polarizations allows the quantification of the

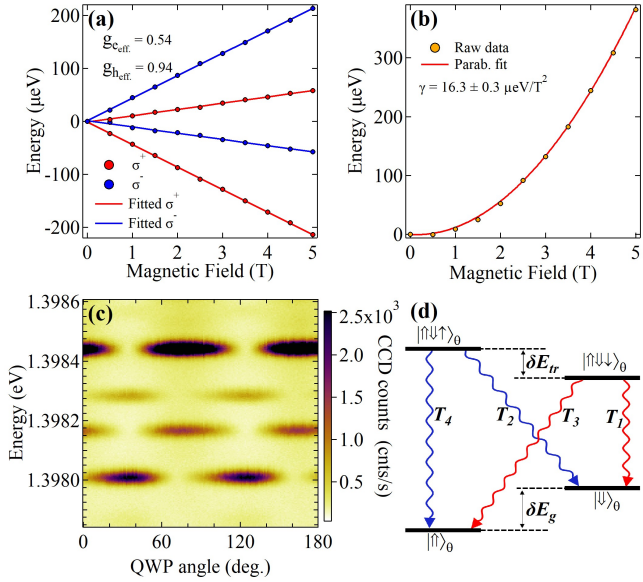
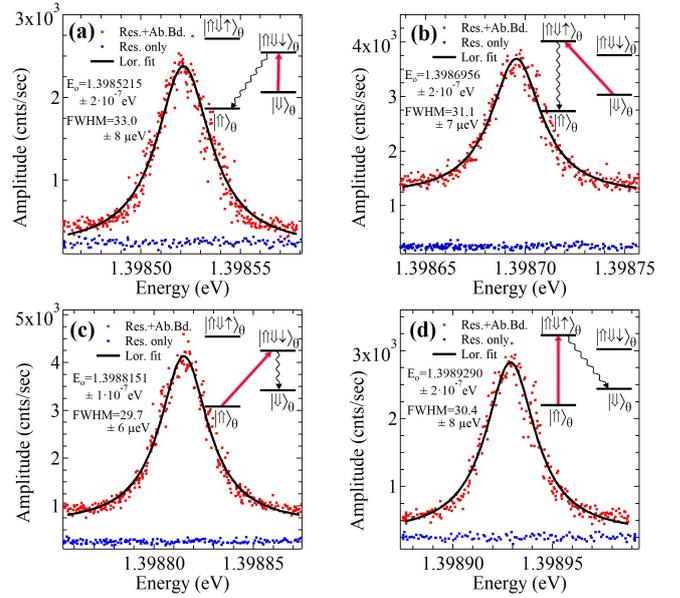

FIG. 2: (a) Progressive splitting of the emission spectra with increasing magnetic field, showing a linear dependence of energy vs magnetic field due to the Zeeman effect. (b) Fitted quadratic dependence due to the diamagnetic shift component. (c) Polarization analysis for each emission peak at 5 T with the $T_1$ & $T_3$ emissions having a right-hand circular polarization, $\sigma^+$, while $T_2$ & $T_4$ have a left-hand circular polarization, $\sigma^-$, the color scale indicates the intensity of each transition in terms of CCD counts. (d) Four-level diagram of the system showing the splitting of the energy levels, where $\delta E_g$ indicates the energy difference between the two ground states and $\delta E_{tr}$ is the difference between the excited states. The red color indicates the $\sigma^+$ polarization while the color blue is for $\sigma^-$.

FIG. 3: (a)-(d) Spin pumping and initialization curves for the four available transitions of the charged QD under 5T magnetic field. The red dots represent the SPCM counts detected from the transition with a wavy downward arrow in the inset, while the resonant laser (solid red arrow in inset) is scanned across the trion transition, with the randomization laser kept on. When the weak above band laser is turned off, the system is initialized to the respective ground hole-spin state and no more photons are observed (blue dots). The solid line is a Lorentzian fit. The insets in the top right corner show the pumping and detection diagram for each transition.

Zeeman splitting from which we find the effective electron and hole g-factors to be $g_e^{eff} = 0.54$ and $g_h^{eff} = 0.94$ as shown in FIG 2(a), and the diamagnetic shift as shown in FIG. 2(b), from which we get the diamagnetic shift factor $\gamma = 16.3 \pm 0.3\,\mu eV/T^2$. The g-factor values that were extracted for this system are very close to previously reported values for similar type of quantum dots [24],[25].

To analyze the polarization characteristics of the four transitions, we perform rotating quarter-waveplate (QWP) polarimetry. We record the emission spectra of the quantum dot for a range of rotation angles of the QWP as shown in FIG. 2(c). Fitting a multi-Lorentzian function to the spectra at each QWP angle, we quantify the polarization properties for each transition. For this quantum dot, all four transitions are predominantly circularly polarized, with polarization degrees exceeding 93%. This aligns with previous findings by Sallen et al. The fourfold energy splitting of the QD and the polarization properties, indicate that we have a four level system with a double-Lambda structure like the one shown in FIG. 2(d).

At this point, it is clear that transitions $T_4$ and $T_2$ have $\sigma^-$ polarization, while $T_3$ and $T_1$ have $\sigma^+$ polarization. The precise understanding of the polarization helps clarify the behavior of the charged exciton states, providing the foundation for the next experiment involving spin pumping in the QD.

### C. Spin-state pumping and initialization

The double-Lambda structure enables us to further investigate this system for its suitability as an optically addressable spin qubit. To this end, we perform spin-pumping and initialization measurements [26]-[35] through all four optically active transitions as shown in FIG.3. This enables us to further verify the double-Lambda structure and to identify the Lambda system to which the two inner transitions belong to. Given that the excitation and detection transitions are on spectrally distinct transitions that are separated by $\sim 250\,\mu eV$, using our high resolution custom built 1m long, double-monochromator with intermediate slits, we can record the detected photon counts with minimal crosstalk from the resonant driving laser.

Low power incoherent excitation from the above-band laser excites carriers at higher energy which through a sequence of non-spin-preserving decay processes, leads to a random population of the excited QD trion levels.

From either of the two trion states, the system can decay into the $|\Uparrow\rangle_\theta$ or $|\Downarrow\rangle_\theta$ ground state, each with a probability that is related to the optical activity of the transition. This

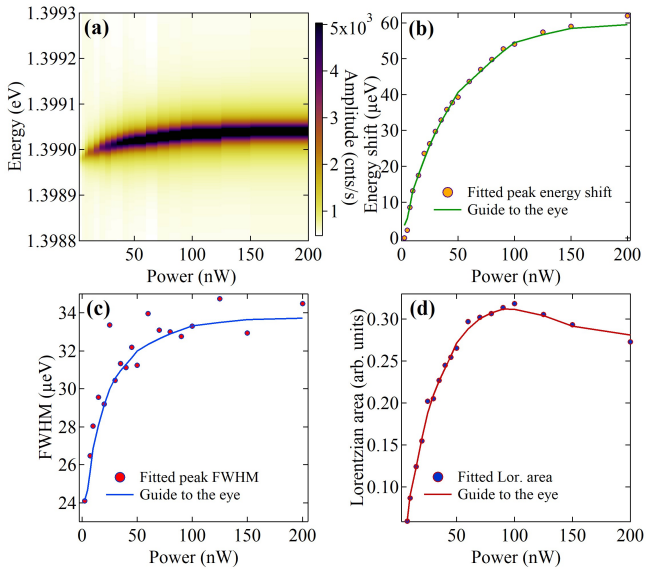

FIG. 4: (a) Lorentzian fits on spin pumping curves for all resonant excitation powers investigated. The pumping configuration investigated here is the same as in FIG. 3(d). (b) Energy dependence of the emission peak with the power of the resonant laser. (c) Spectral width evolution with increasing resonant power. (d) Area of the emission peaks as a function of the resonant laser power.

effectively allows to randomize the ground spin states. For this process we applied $P_{Ab.Bd.} = 10\,\text{nW}$ which corresponds to 4% of the saturation power.

For the resonant laser, the power was set to $P_{Res.} = 15\,\text{nW}$. By fine-tuning the resonant laser frequency in resonance with one of the two branches of a Lambda system while keeping the above-band laser active, we observe a Lorentzian-shaped peak in the detected counts coming from the other branch of the Lambda system. For instance, when the resonant laser is on transition $T_1$, we detect photons from transition $T_3$ (diagram on inset of FIG. 3(a)). This happens because the resonant laser excites the system to the $|\Uparrow\Downarrow\downarrow\rangle_\theta$ state which through spontaneous emission eventually decays to $|\Uparrow\rangle_\theta$ giving out a photon at the energy of the transition $T_3$. As the above-band laser helps randomize the ground states, the exact same process happens continuously, giving rise to the red dot spin-pumping peak observed in FIG. 3(a). Turning off the weak above-band laser, while maintain the resonant laser in resonance causes a drop in photon counts to the background level, even with the resonant laser still fine-tuned in resonance to the $T_1$ transition frequency as shown in FIG. 3(a) blue-dot trace. This occurs because without the above-band laser, once the system is initialized to the $|\Uparrow\rangle_\theta$ state, the opposite ground state is empty and the resonant laser cannot drive the transition $T_1$ until the next randomization event. In this case, the SPCM detects only background counts.

The lack of additional peaks in the background data confirms high-fidelity optical initialization of the spin in the $|\Uparrow\rangle_\theta$ state. The same spin pumping and initialization experiment can be performed on all three remaining transitions by exciting $T_2$ and detecting $T_4$ (FIG. 3(b)), exciting $T_3$ and observing $T_1$ (FIG. 3(c)), and exciting $T_4$ while detecting $T_2$ (FIG. 3(d)). Each case shows a clear spin-pumping peak and the same level of high-fidelity initialization, highlighting the potential for all-optical spin-state control across all four transitions.

### D. Resonant power dependence

In the final experiment, we investigate the effect of varying the resonant laser power on the driven QD transition while maintaining the same low power for the above-band excitation as before. Here, the resonant laser is brought in resonance with the highest energy transition $T_4$, while detecting photons from $T_2$, akin to the experiment in FIG. 3(d) and the power is increased in the range $P_{Res.} = 2.5 - 200\,\text{nW}$ over 18 steps. For each power, the spin pumping curve is fitted with a Lorentzian and the experiment is repeated with the above-band laser off, in order to estimate the initialization fidelity. The Lorentzian fits for all investigated powers are shown in FIG. 4(a) from which one can see a combination of blue-shifting and broadening taking place. Using the information extracted from the fit, we plot the Lorentzian peak energy as a function of resonant laser power in FIG. 4(b). The peak energy exhibits a significant blue-shift, reaching about 60 µeV at the highest powers. FIG. 4(b) shows the Full Width at Half Max (FWHM) of the Lorentzians, indicating a spectral broadening, with then FWHM increasing by approximately 30% over the full power range. This is likely due to power-induced dephasing/broadening effects. Additionally, we quantify the spin pumping peak intensity using the area of the fitted curves which initially increases up to up to a threshold of ~100 nW, beyond which it slightly decreases, suggesting emission saturation due to state depletion or heating effects, as shown in FIG. 4(d). Interestingly the initialization fidelity remains unchanged throughout the complete range of resonant laser powers used.

## IV. Conclusions

In this study we investigate spin pumping and initialization of an individual hole-spin in a site-controlled pyramidal quantum dot under a magnetic field in the Faraday configuration. Thanks to the high symmetry of these quantum dots application of such a magnetic field creates a four-level system with spin and optical properties resembling those of Stranski-Krastanov QDs in oblique magnetic fields. This gives rise to an optically accessible double-lambda level structure with predominantly circularly polarized transitions which we here employ for spin-pumping and high-fidelity spin-initialization under a broad variety of driving configurations and conditions.

These results demonstrate that site-controlled pyramidal quantum dots in (111)B GaAs are promising candidates for advanced spin-based quantum applications and pave the way towards experiments demonstrating universal single qubit gate operation in this highly versatile and scalable QD platform.

## ACKNOWLEDGEMENTS

R.A.B. acknowledges financial support from the EPSRC Doctoral Training Program under grant no. EP/Y011864/1 and the Agency for Student Loans and Scholarships (https://roburse.ro) under the scholarship H.G. no. 118/2023.

I.S. acknowledges financial support from the EPSRC Doctoral Training Partnership under grant no. EP/W524670/1. K.B. acknowledges financial support from the EPSRC Doctoral Training Program under grant no. EP/R513349/1. G.J. and E.P. acknowledge funding from Research Ireland, formerly Science Foundation Ireland, under Grants Nos. 22/FFP-P/11530, 22/FFP-A/10930, 15/IA/286